\documentclass[pra,twocolumn,showpacs,floatfix,amsmath,amsfonts]{revtex4}

\newcommand{\cU}{{\cal U}}
\newcommand{\cG}{{\cal G}}
\newcommand{\cE}{{\cal E}}
\newcommand{\cF}{{\cal F}}

\usepackage{epsfig}
\usepackage{graphicx}
\usepackage{amsmath}
\usepackage{amssymb}
\usepackage{dcolumn}

\begin{document}
 \title{Rabi oscillations of matter wave solitons in optical lattices}

\author{Yu. V. Bludov$^{1}$}

\author{V. V. Konotop$^{2}$}

\author{M. Salerno$^{3}$}

\affiliation{$^1$ Centro de F\'{\i}sica, Universiade do Minho, Campus de Gualtar, Braga 4710-057, Portugal
\\
$^2$Centro de F\'{\i}sica Te\'orica e Computacional,
Universidade de Lisboa, Complexo Interdisciplinar, \\ Avenida
Professor Gama Pinto 2, Lisboa 1649-003, and Departamento de F\'{\i}sica,\\ Faculdade de Ci\^encias,
Universidade de Lisboa, Campo Grande, Ed. C8, Piso 6, Lisboa
1749-016, Portugal\\ $^3$Dipartimento di Fisica "E. R. Caianiello"
and Consorzio Nazionale Interuniversitario  per le Scienze\\
Fisiche  della Materia (CNISM),  Universit\'a di Salerno, via S.
Allende I-84081, Baronissi (SA), Italy}

\begin{abstract}

Inter-band  Rabi oscillations of
gap soliton matter waves  induced by time dependent
periodic forces in combined linear and nonlinear optical
lattices are for the first time demonstrated.
It is shown that under suitable conditions these oscillations
can become long-lived. By switching off the external force at proper time
it is possible to create either
pure (stationary macroscopically populated) gap
soliton states or linear combination
of two gap solitons with  appreciably long
life-time.
\end{abstract}
 \maketitle

\section{Introduction}

In spite of its long history, the phenomenon of quantum tunneling still
attracts a great deal of attention, particularly in connection with
its macro-world manifestations. In condensed matter
physics tunneling phenomena occur in connection with inter-band transitions
of electrons (from  the valence to the conduction band)
under the action of external perturbations such as electric fields
(Landau-Zener tunneling \cite{Zener}) or radiation fields (photons) of proper energy
(inter-band Rabi oscillations). The phenomenon is highly
facilitated by time periodic perturbations with
frequencies  matching  the energy gaps between the bands.
In the context of cold atoms Rabi oscillations (RO) between Bloch bands have been
considered  both theoretically \cite{niu1} and
experimentally \cite{niu2}. Possible
extensions of these studies to ultracold atoms and to Bose-Einstein condensates (BEC)
give  an interesting  perspective for inter-band transitions
and ROs detection at the macroscopic level.

Macroscopic quantum tunneling phenomena are presently investigated in many
physical systems,  including superfluidity and superconductivity \cite{Takagi} and,  more
recently, in optical systems with ultra-cold atoms~\cite{atoms}
and Bose-Einstein condensates (BECs) embedded in optical lattices
(OLs)~\cite{MO}, where the Landau-
Zener tunneling has been recently observed~\cite{tunnel_BEC}. In contrast with
solids, however, BEC systems are intrinsically nonlinear, being
described by the mean field periodic Gross-Pitaevskii equation
with the cubic nonlinearity arising from the interatomic two-body
interactions. The correspondence with solid state physics occurs
either at very low densities or when the interatomic interaction
are "artificially" detuned to zero  by means of external magnetic
fields, using Feshbach resonances (see e.g.~\cite{Feshbach}).

The presence of nonlinearity in the periodic Gross-Pitaevskii
equation introduces modulational (dynamical) instabilities of
Bloch wavefunctions \cite{Inguscio} with substantial effects
on the inter-band tunneling~\cite{KKS,tunnel_BEC}. The appearance
of spatially localized states inside gaps of the underlying linear
band is associated to these modulational instabilities. These
states, also known as gap-solitons (GSs) first discovered in
nonlinear optics~\cite{GS} and observed in fibers with Bragg
gratings~\cite{BrG}, in BECs were experimentally realized for the
first time in~\cite{GS-BEC}. Such states have energies which can
scan the whole gap (or a large portion of it) giving rise to
families of modes with different symmetry properties with respect
to the OL (see e.g.~\cite{BK}).
The
existence of GSs with chemical potentials arbitrarily close to
band edges
introduces
new possibilities for  a stimulated Landau-Zener tunneling and RO of matter waves
across the gap.

The aim of the present paper is to show for the first time how
the inter-band {\it Rabi oscillations} of matter wave solitons
can be induced by means of combined linear
and nonlinear OLs under the action of a time periodic linear potential.
By switching off the external force at proper times we show that
it is possible to put the system either in a  stationary GS state
or in a time dependent linear combination of two
GS states which, in spite of the nonlinearity of the system, can
persist in this superposition for a reasonable long time.

The paper is organized as follows. In section II we introduce the model equations
while in section III  we provide an analytical description of the long-lived Rabi oscillations  of GS
in terms of a modified two-mode model valid for GS states close to
gap edges. A small detuning of the force frequency from the gap-width will be
used as parameter to achieve the absence of matter leakage
(emission of radiation) from the soliton states.
We show that under this condition the ROs become long lived and manageable.
In Section IV we compare our analytical results with direct numerical simulations of the Gross-Pitaevskii equation.
In this section we also  discuss the possibility  to create pure (stationary macroscopically populated) gap
soliton states or linear combination
of two of them which survive for appreciably long time, by switching off the external force
sustaining the Rabi oscillation at a proper time.
In the last section we  design parameters for a possible experimental observation of the nonlinear ROs  and
briefly summarize the main results of the paper.

\section{Model equation}

Although in the following we refer specifically to a BEC, the obtained results will be valid for
all physical systems modeled by the following periodic nonlinear
Schr\"{o}dinger (NLS) equation with time dependent linear potential:
\begin{eqnarray}
\label{psi_fin} i\psi_t=-\psi_{xx}+\cU(x)\psi+\cF(t)\;
x\psi+\cG(x)|\psi|^2\psi,
\end{eqnarray}
Here $\cU(x)$ and $\cG(x)$ are $\pi$-periodic linear and
nonlinear lattices, respectively, and $\cF(t)$ is the time periodic
amplitude of a linear potential.
In BEC experiments the nonlinear lattice can be created by
optically~\cite{Shliap} or magnetically~\cite{magnetic} induced
Feshbach resonances (for the sake of brevity we speak about
OL) while the external force can be induced simply by a time
periodic acceleration of the linear OL.
In absence of the nonlinearity, $\cG(x)\equiv 0$ and of the
linear potential $\cF(t)\equiv 0$, Eq. (\ref{psi_fin}) reduces to the
familiar band theory framework, with the Schr\"odinger
operator ${\cal H}=-d^2/dx^2+\cU(x)$ and associated stationary
eigenvalue problem ${\cal
H}\varphi_{nq}(x)=\cE_n(q)\varphi_{nq}(x)$, this providing
energy bands $\cE_n(q)$ and Bloch functions $\varphi_{nq}(x)$
with quasi-momentum  $q$ and band index $n$, normalized as
$\int_{-\pi}^{\pi}|\varphi_{nq}(x)|^2dx=1$.
We will concentrate on the lowest gap extending
from the top of the first band, denoted as $\cE_-$, to the bottom
of the second band, denoted as $\cE_+$. The respective Bloch
functions will be denoted by $\varphi_\pm(x).$
\begin{figure}[tb]
       \scalebox{1.0}[.9]{\includegraphics{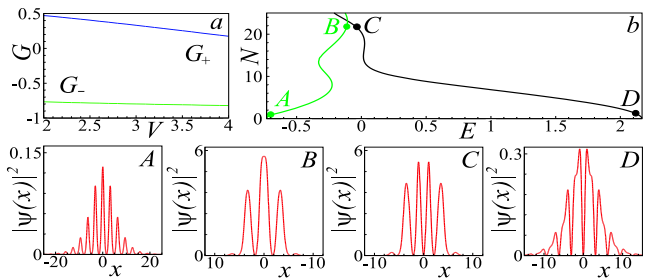}}
\caption{(a) Curves  $G_\pm(V)$ separating domains where
$\chi_\pm(V,G)>0$ and $\chi_\pm(V,G)<0$ (above and below the
respective lines) with   $+$ ($-$) denoting
quantities evaluated at the top (bottom) edge of first
gap. (b) Soliton norm $N$ {\it vs} energy $E$  of lowest
symmetric (left curve) and antisymmetric (right curve) GS in the
gap $\cE_-<E<\cE_+$ (the edges of the gap
coincide with the  boundaries of the $E$-axis) for the parameters  $V=3$, $G=-0.5$. Lower panels show
shapes of the localized modes at points A -- D of the panel b. }
\label{fig:chi}
\end{figure}

The condition for the existence of GS states at
both edges of a gap can be expressed in terms of effective masses
$M_ \pm =\left(d^2\cE_\pm /dq^2\right)^{-1}$ and effective
nonlinearities $\chi_ \pm =\int_{-\pi}^{\pi}\cG(x)|\varphi_\pm
(x)|^4dx$,  as: $M_ \pm \chi_\pm<0$~\cite{SKB}. In analogy to what
done in Refs. \cite{BLK,SKB}, we satisfy this condition by
considering OLs of the form: $\cU(x)=-V\cos(2x)$,
$\cG(x)=G+\cos(2x)$, with $V$ and $G$ related to the physical amplitude of the linear OL ${\cal V}$ and to the scattering length $a_s(x)=a_{s0}(G+\cos(2x))$
by ${\cal V}=V E_R$ and  $G=\langle a_{s}\rangle /a_{s0}$ where $\langle a_s\rangle=\pi^{-1}\int_{0}^{\pi}a_s(x)dx$ denotes the average of the scattering length over one period of the nonlinear OL. Here $E_R={\hbar^2\pi^2}/(2 m d^2)$ is the recoil energy, $d$ is the lattice constant,  $a_{s0}>0$ is the amplitude of the scattering length modulation, and  $m$ the mass of an atom. Space and time
variables in our formulas are measured in units $d/\pi$,
and $\hbar/E_R$, respectively, with the norm $N$ related to the physical number of atoms ${\cal N}$  by $N=\frac{4 a_{s0} d}{\pi a_\bot^2}{\cal N}$ and with $a_\bot$ the transverse oscillator length.
From Fig.~\ref{fig:chi}a we see
that the OLs appropriate for our task are the ones with $G,V$
taken in the region between curves $G_\pm$. Notice that for fixed $G$ and $V$ a family of solutions, parameterized by the soliton norm  $N=\int_{-\infty}^\infty|\psi|^2dx$  (for a cigar-shaped configuration   bifurcates from the opposite band edges (panel b). The
bottom panels of Fig.~\ref{fig:chi} illustrate change of the symmetry  of a GS  as the different edges of the gap.

To stimulate RO between GSs
we use a force $\cF = \nu \cos (\omega t)$ with
amplitude $\nu \ll 1$ and frequency $\omega=\cE_g-2\Delta$
slightly detuned from the gap width $\cE_g=\cE_+-\cE_-$ (i.e.
$\Delta \ll \cE_g$).
The oscillation must involve a nontrivial rearrangement of the
matter at each cycle, due to the change of symmetry of
the state and the corresponding dynamics can be sustained for a
long times only if the force is properly designed. In the
following we use $\Delta$ as a free parameter to find
optimal inter-band tunneling conditions.

\section{Two mode model for nonlinear Rabi oscillations}

\begin{figure}
  \begin{center}
   \begin{tabular}{c}
       \scalebox{1.}[.9]{\includegraphics{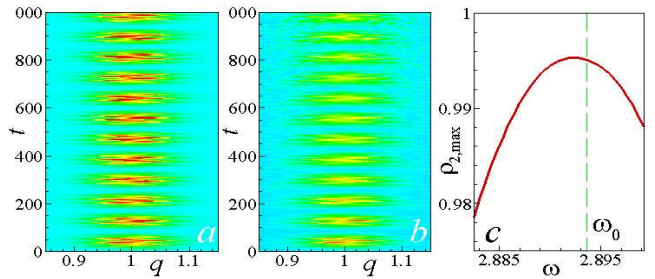} }
   \end{tabular}
   \end{center}
\caption{Long-lived (a) and fast decaying (b) Rabi oscillations in
momentum space $|c_2(q)|^2$.
(c) Maximal population of the
second band as obtained from NLS numerical integrations
in the interval  $\omega \in \left[\cE_g - (\Omega_- - \Omega_+), \cE_g\right]$
around the optimal value $\omega_0$ (dashed line).
Other parameters are $\nu=0.1$, $V=3$, $G=-0.644$ (panels a,c], and
$G=0.4$ (panel b). The energy of the initial state is $E=0.73$.
}
\label{fig:opt-nonopt}
\end{figure}

An analytical description of the phenomenon can be made in terms
of a two-level  model.
To this end we consider  $\sqrt{\nu/\omega}\ll 1$  as a small
parameter and recall the semi-classical equation of motion
$dq/dt=-\nu\cos(\omega t)$,  leading to the change of the phase of
the wavefunction as follows $\displaystyle{e^{i [q(t)- q(0)]
x}=e^{-i(\nu x/\omega)\sin(\omega t)}}$, and take into account
that in the leading order the solution is a superposition of the
states bordering the gap edges:
\begin{equation}
\displaystyle{\psi\approx  A_+ \varphi_+ e^{-i(\cE_+-\Delta) t}+ A_-  \varphi_- e^{-i (\cE_-+\Delta)t} }.
\label{ansatz}
\end{equation}
Here $A_\pm\equiv A_\pm (x,t)\sim\sqrt{\nu/\omega}$ are  slow
varying  amplitudes  ($\partial A_\pm/\partial t\sim \partial^2
A_\pm/\partial x^2\sim (\nu/\omega) A_\pm$) of the Bloch states
$\varphi_\pm$ bordering the edges with energies $\mu_\pm= \cE_\pm
\mp \Delta$, providing $\mu_+ - \mu_- = \omega$ and satisfying the above criteria
$\Delta\sim\nu\ll\omega$ being small detuning towards the gap.
Dropping details (see e.g.~\cite{KKS,BK}), we give the final
system for the evolution of the amplitudes
\begin{eqnarray}
i\frac{\partial A_\pm}{\partial t} = &-&\frac{1}{2M_\pm }
\frac{\partial^{2} A_\pm}{\partial x^{2}} \pm \Delta A_\pm
+\gamma A_\mp \nonumber \\
&+& (\chi_ \pm  |A_\pm|^{2}+2\chi  |A_\mp|^{2})A_\pm,
\label{eq:apm}
\end{eqnarray}
where $\chi =\int_{-\pi}^{\pi}\cG(x ) \varphi_+^2 \varphi_-^2 dx $
and $ \gamma = \frac{\nu}{\omega}
\int_{-\pi}^{\pi}\frac{d\varphi_+}{dx} \varphi_-  dx $.
In absence of nonlinearity and for
spatially homogeneous amplitudes this system reduces  to the model
of RO of two-level atoms in a light
field~\cite{Rabi_atom}. This confirms the interpretation
of the phenomenon as {\em Rabi oscillations of a gap
soliton} between the two energy levels $\mu_\pm$.
Neglecting the second derivative and nonlinearity in Eq.(\ref{eq:apm}),
the inter-band tunneling time can be estimated as half
of the Rabi period, i.e.  $T_{tun}=\pi/\Omega$,  with $\Omega$
the usual linear Rabi frequency $\Omega=2\sqrt{\gamma^2+\Delta^2}$.
By approximating Bloch function at the first gap edges as
$\varphi_+(x)\approx \pi^{-1/2}\sin(x)$, $\varphi_-(x)\approx
\pi^{-1/2}\cos(x)$, we readily estimate $T_{tun}\approx
\pi/(2\sqrt{\nu^2/\omega^2+\Delta^2})$.
\begin{figure}
  \begin{center}
   \begin{tabular}{c}
       \scalebox{1.0}[.9]{\includegraphics{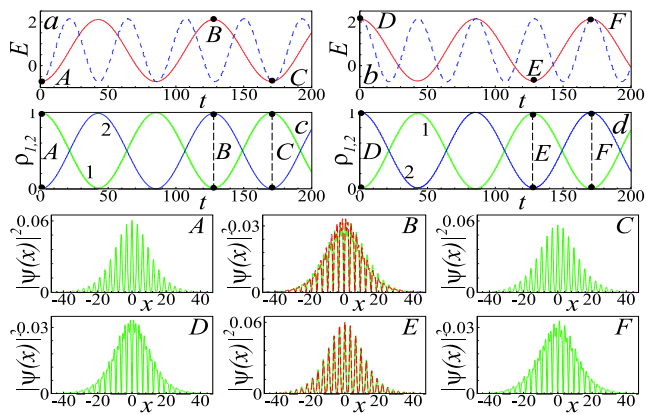}}
   \end{tabular}
   \end{center}
\caption{(a,b) $E$ versus time for $\nu=0.1$ (solid lines)
and $\nu=0.2$ (dashed lines);  (c,d) dynamics of the populations,
$\rho_{1,2}(t)$, of the two bands for $\nu=0.1$. Parameters
are: $V=3$, $G=-0.644$ and the initial conditions correspond to
$E=-0.73$ (panels a,c) and $E=2.152$ (panels b,d). Lower panels
show the solitonic shapes corresponding to points A--F. In
panels B and E soliton profiles are compared to stationary
solutions (dashed lines) with the same $N$ of the initial conditions ($N=0.61$), but with
energies in the vicinity of the opposite gap edge ($E=2.152$ in
panel B and $E=-0.73$ in panel E).} \label{fig:nt}
\end{figure}
The two GS states at gap edges can be estimated
from the  soliton solutions of Eq. (\ref{eq:apm}) given by
$ A_\pm^{(s)}=0,\;
A_\mp^{(s)}=\sqrt{\frac{2\Omega_\mp}{\chi_\mp}} e^{ - i(\Omega_\mp\mp
\Delta) t } \mbox{sech}[\sqrt{-2M_\mp\Omega_\mp}x]$, with  $M_\pm\Omega_\pm<0$
and $\Omega_\mp$ fixed by
$N=\sqrt{2 |M_{\mp} \Omega_{\mp}|}/\pi|M_{\mp} \chi_{\mp}|$. In the following
we use this solution to find an optimal design for long-lived ROs.
To this regard, we remark  that long-lived ROs can be achieved
only if a minimal loss  of matter (ideally zero) from the solitons occurs
during each oscillation cycle. For this one must impose  the conservation of the energy
and the conservation of $N$ (complete exchange of particles between components
at the turning points). Moreover, since the states must change  symmetry every half cycles
(being states at opposite gap edges) one must  also  require that the spatial localization of the two
GSs at the turning points is approximately the same (this facilitates  the
symmetry exchange). The last  condition is satisfied
if $M_+\chi_+ = M_-\chi_-$, while the conservation
of $N$ and  the conservation of the energy imply $M_-\Omega_-=M_+\Omega_+$
and  $\Delta=\left(\Omega_- - \Omega_+ \right)/6$, respectively.
Since $\Omega_\pm$ depends on $N$, $M_\pm$ ,$\chi_\pm$,  one
can adjust these parameters to have all the above relations satisfied with the
driving frequency $\omega_0=\cE_g-2 \Delta$ chosen as
$\omega_0=\cE_g-(\Omega_--\Omega_+)/3$ for an {\it optimal} design
(see Fig.\ref{fig:opt-nonopt}c). Notice that for $\omega \ne \omega_0$,  while  $N$ will be conserved,
the soliton energy will not be perfectly  conserved during the tunneling process and that for fixed $M_\pm$ ,$\chi_\pm$, a change
of $N$ requires the adjustment of the optimal frequency. Contrary to the linear case, where
the optimal transfer between levels is achieved for zero detuning,
here the optimal transfer is obtained for $\Delta\neq0$. This is a
direct consequence of the nonlinearity of the system which replaces
linear atomic wavefunctions with energies exactly at gap edges
with nonspreading GSs wavepackets with energies
detuned from band edges inside the gap. In the limit of zero nonlinearity ($N\to 0$) GSs
tend to  extended Bloch states at gap edges and the linear optimal condition ($\Delta=0$)
for RO is recovered.

\section{Numerical results}

In order to check these predictions we have performed direct numerical
integrations of Eq.(\ref{psi_fin}) with  initial condition in a
form of a stationary GS close to the bottom edge of the
gap. The frequency detuning, the parameter values of the OLs, and
the initial conditions were chosen to met the requirements of
conservation of energy, soliton norm and widths in the
solitonic states during the oscillation, as described above.

\subsection{Nonlinear Rabi scillations and tunneling time}

The results are depicted in Fig.\ref{fig:nt} where, in panels a,b,
we show that minima and maxima of the periodically varying energy
of the system well match the GS levels, what represents the
most clear evidence of the inter-band tunneling.
Notice that the frequency of the energy oscillations increases with the force strength $\nu$ and correlates
with similar oscillations of the band populations
(Fig.~\ref{fig:nt}c,d) computed as
$\rho_{1,2}=\int_{-1}^{1}|c_{1,2}(q)|^2 dq/N$, where
$c_n(q)=\int_{-\infty}^{\infty}\psi \overline{\phi}_{n,q} dx$.
One also observes change  of the wave-packet symmetry:
initially symmetric (with respect to $x=0$) soliton excited in the
vicinity the lower gap edge (Fig.~\ref{fig:nt} A), in the
vicinity of the upper gap edge transforms into an odd solution
(Fig.~\ref{fig:nt} B). Similarly, initial odd soliton
(Fig.~\ref{fig:nt} D) is transformed into the even one
(Fig.~\ref{fig:nt} E). After one cycle of oscillations
(points C and F in the upper panels of Fig.~\ref{fig:nt}), the
soliton approximately restores its initial shape (c.f. panels A
and C, and panels D and F).
\begin{figure}
  \begin{center}
   \begin{tabular}{c}
       \scalebox{1.0}[.9]{\includegraphics{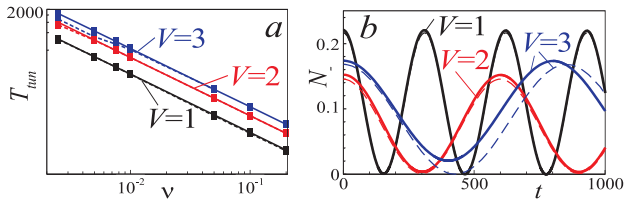}}
   \end{tabular}
   \end{center}
\caption{  (a)  $T_{tun}$ {\it vs}  $\nu$ (in the
logarithmic scales);
(b) evolution of the soliton norm in the lowest band $N_-$ at $\nu=0.01$,  obtained from the direct numerical simulation of
(\ref{psi_fin}) (solid lines) and from the two-mode approximation
(\ref{eq:apm}) (dashed lines).
}
\label{fig:ttun-V}
\end{figure}

To compare RO under optimal and arbitrary
conditions, we show  in Fig.~\ref{fig:opt-nonopt}a,b, the time evolution of the GS states
projected on the second band (i.e. the population
$|c_2(q)|^2$)
as obtained from numerical integrations of the NLS  in the case of an optimally chosen nonlinearity $G$
(Fig.~\ref{fig:opt-nonopt}a)  and for a non optimal value  of $G$ for which
small-amplitude GSs do not exist close to the upper band edge (Fig.~\ref{fig:opt-nonopt}b).
We see that while in the first case the ROs are long-lived, in the latter  case the soliton is destroyed
after few periods of oscillation. From Fig.~\ref{fig:opt-nonopt}c we
also see that the numerical frequency for the maximal transfer  of atoms from one band to another
is in very good agreement with the analytical value  $\omega_0$ estimated from the two mode model.

In  Fig.~\ref{fig:ttun-V} we present the comparison of the tunneling
time $T_{tun}$  obtained from the direct numerical simulations and
in  the framework of the two-mode model. One observes that $T_{tun}$ decreases with the increasing the  force
strength $\nu$, and increases with the amplitude of the linear OL.
The discrepancies between the exact and approximated models are
negligible for moderate amplitudes of the linear OL (namely
$V\lesssim 2$ in Fig.~\ref{fig:ttun-V}a), and become appreciable
for a deep OL and small external force strength ($V=3$ and
$\nu\lesssim 0.01$ in Fig.~\ref{fig:ttun-V}a and $V=3$ in
Fig.~\ref{fig:ttun-V}b).

\begin{figure}
  \begin{center}
   \begin{tabular}{c}
       \scalebox{1.0}[1.0]{\includegraphics{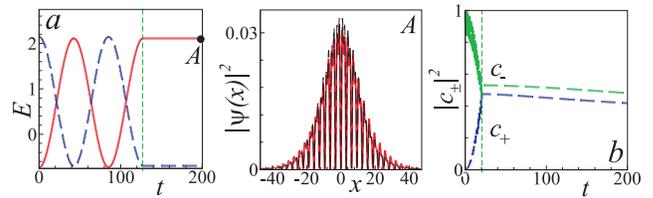}}
   \end{tabular}
   \end{center}
\caption{(a) The soliton energy $E(t)$ for
 $E(0)=-0.73$ (solid lines) and $E(0)=2.152$ (dashed
lines), with the linear potential switched off at
$T_{tun}\approx 42.6$ (the vertical dashed
line). Parameters of OLs are fixed as in
Fig.~\ref{fig:nt} but with $\nu=0.1$. Panel A shows the soliton shapes at
  point A of panel a (solid lines), compared with the
 stationary solutions (dashed lines) depicted in
Fig.~\ref{fig:nt}B.  (b) Population densities $| c_{\pm}(t) |^2$
after the switching off of the linear potential at $t=20.6$.
}
\label{fig:ntoff}
\end{figure}

\subsection{Superposition of two gap soliton states}

It is worth to note that by switching off the linear potential at a time
$t_s=(2n+1)T_{tun}$ ($n=0,1,...$) the state of the system will
coincide with  one of the stationary states close to a band edge (see Fig.~\ref{fig:ntoff}a and panel A in the same figure).
On the other hand, from Eq. (\ref{ansatz}) we see that by switching off the
linear potential at any other time $t_s\ne (2n+1)T_{tun}$ will produce
a state which is a linear superposition of the two stationary
states at the edges of the gap:  $\psi_{s}= c_{-}(t)  \psi_{-}  + c_{+}(t) \psi_{+} $,
where $ c_{\pm}= \frac 1N \int \psi(x,t) \psi_{\pm}(x) dx$ are complex amplitudes and
$\psi_{\pm}$ are the numerically exact GSs with the same $N$
used in direct simulations.
It is remarkable that, in spite of the nonlinearity of the system, the decay rate of the
superposition state (slopes of  $c_{\pm}$) may be small enough for the state to be
observed. The decay rate depends, however, on the soliton norm, and
we find that for $N=0.17$ is $\sim 4.5\cdot 10^{-5}$, at  $N=0.61$ is $\sim 3.5\cdot 10^{-4}$ (see Fig.~\ref{fig:ntoff}c) and for $N=1.24$  it becomes $\sim 1.6\cdot 10^{-3}$,
as consequence of  the increased effective nonlinearity in the system.

\section{Discussion and conclusions}

We discuss a possible parameter design for  experimental observation
of long-lived Rabi oscillations between  gap soliton states at
opposite edges of a band gap. Referring to a $^7Li$ condensate in a trap  with $a_\bot=2\, \mu {\rm m}$,
$d=1\,\mu {\rm m}$  and with $\langle a_{s}\rangle=-1.288\,{\rm
nm}$, $a_{s0}=2\,{\rm nm}$ created by an optically induced Feshbach
resonance, the RO depicted in Figs.~\ref{fig:nt}c,d can be achieved by an external force of
$1.45\cdot 10^{-24}\,{\rm N}$ (corresponding to an acceleration of
the OLs $\approx 124\,{\rm m/s}^2$), oscillating with the
frequency $20.38\,{\rm kHz}$ and applied to a GS with about ${\cal N}=980$ atoms (this corresponding to $N=0.61$).
The tunneling time and the decay rate of the superposition state in Fig.~\ref{fig:ntoff}c are estimated
as $0.965\,{\rm ms}$  and $\sim 15.5\,$s$^{-1}$, respectively.

In summary, we have shown for the first time how to achieve long-lived inter-band Rabi oscillations of gap-solitons
matter waves  in combined linear and nonlinear optical lattices. We also showed that by switching off the external
force at proper time it is possible to create either pure (stationary macroscopically populated) gap
soliton states or linear combination of two gap solitons with  appreciably long
life-time.

We finally remark that similar phenomena are expected to occur also in arrays of optical waveguides periodically
modulated along the
propagation direction.

\section{Acknowledgements}

YVB was partially supported by the FCT Grant No. SFRH/PD/20292/2004. Partial support from a cooperative bilateral agreement between FCT (Portugal) and
CNR (Italy) is acknowledged.

\end{document}